\documentclass[aps,prb,reprint,groupedaddress,showpacs,amsmath,amssymb]{revtex4-1}
\usepackage{graphicx}
\usepackage{bm}

\begin{document}

% Use the \preprint command to place your local institutional report
% number in the upper righthand corner of the title page in preprint mode.
% Multiple \preprint commands are allowed.
% Use the 'preprintnumbers' class option to override journal defaults
% to display numbers if necessary
%\preprint{}

%Title of paper
\title{Scaling analysis of field-induced superconductor-insulator transition in two-dimensional tantalum thin films}

% repeat the \author .. \affiliation  etc. as needed
% \email, \thanks, \homepage, \altaffiliation all apply to the current
% author. Explanatory text should go in the []'s, actual e-mail
% address or url should go in the {}'s for \email and \homepage.
% Please use the appropriate macro foreach each type of information

% \affiliation command applies to all authors since the last
% \affiliation command. The \affiliation command should follow the
% other information
% \affiliation can be followed by \email, \homepage, \thanks as well.

\author{Sun-gyu Park}
\affiliation{Center for Supersolid and Quantum Matter Research and Department of Physics, KAIST, Daejeon 305-701, Republic of Korea}

\author{Eunseong Kim}
\affiliation{Center for Supersolid and Quantum Matter Research and Department of Physics, KAIST, Daejeon 305-701, Republic of Korea}
\email{eunseong@kaist.edu}

%Collaboration name if desired (requires use of superscriptaddress
%option in \documentclass). \noaffiliation is required (may also be
%used with the \author command).
%\collaboration can be followed by \email, \homepage, \thanks as well.
%\collaboration{}
%\noaffiliation

\date{\today}

\begin{abstract}
The superconductor-insulator (SI) transition in two-dimensional Ta thin films is investigated by controlling both film thickness and magnetic field. An intriguing metallic phase intervening superconductor and an insulator phase is observed within a narrow range of film thickness and magnetic field. Finite scaling analysis has been performed to investigate the nature of the SI transition in the thickness-tuned metallic and superconducting samples. The critical exponents in the disorder-induced metallic samples are clearly different from the exponents obtained in the superconducting samples. Dynamical exponent \textit{z} of the superconducting samples is consistent with the theoretical predictions (\textit{z} = 1), while the exponent for the metallic samples is approximately 0.7. The discrepancy in the transition behaviors supports that the disorder induced metallic phase cannot be classified to the same universality class of the superconducting Ta thin films.
\end{abstract}

% insert suggested PACS numbers in braces on next line
\pacs{74.40.Kb, 74.78.-w, 74.25.F-,64.70.Tg}
% insert suggested keywords - APS authors don't need to do this
%\keywords{}

%\maketitle must follow title, authors, abstract, \pacs, and \keywords
\maketitle

% body of paper here - Use proper section commands
% References should be done using the \cite, \ref, and \label commands
% Put \label in argument of \section for cross-referencing
%\section{\label{}}
%\subsection{}
%\subsubsection{}

The quantum phase transition\cite{Sondhi1997} of a two-dimensional disordered superconducting system can be achieved by amplitude fluctuation\cite{Hsu1995} or phase fluctuation\cite{Hebard1990,Yazdani1995,Markovic1998} of superconducting order parameters. According to the ‘dirty boson’ model,\cite{Markovic1998} the SI transition can be obtained by phase fluctuations induced by increasing either the disorder or the magnetic field. Within this framework, a superconducting phase corresponds to a vortex glass state in which a condensate of Cooper pairs appears with localized vortices, while an insulating phase corresponds to a Bose glass state in which Cooper pairs are localized at the proliferated vortices. Scaling analysis demonstrates the existence of the universal critical resistance and also indicates that the critical exponents are $\nu \geq$ 1 and \textit{z} = 1 at the critical point.\cite{Fisher1990,Fisher1990a}

According to the scaling theory of localization, a metallic state in two dimensions should be impossible.\cite{Abrahams1979} Because defects localize all types of carriers – either electrons or Cooper pairs – in an infinitely large two-dimensional system at temperature \textit{T} = 0, the system resistance becomes infinite. Accordingly, the two-dimensional quantum phase transition tuned by either the magnetic field or the disorder is expected to be a SI transition.

Recently, Ta\cite{Qin2006,Seo2006,Li2010} and MoGe\cite{Ephron1996,Mason1999} thin films show the magnetic field induced unexpected intermediate phase between superconducting and insulating phase in a wide range of magnetic field. This intermediate phase is considered as a metallic phase because of the finite saturated resistance at low temperatures. Possible mechanisms such as quantum vortex metal and quantum phase fluctuation for the unusual behaviors were proposed.\cite{Das1999,Das2001,Spivak2001,Dalidovich2002,Galitski2005,Kapitulnik2001} However, the underlying physics of this unexpected phase is not fully understood. In addition to the magnetic field induced metallic phase, Ta thin films with a certain range of thickness exhibit a disorder induced metallic phase\cite{Li2010} even at the zero field and zero temperature limit. The IV characteristics\cite{Qin2006} of metallic phase distinct from the other phases are also reported.

The scaling behavior of two-dimensional disordered superconducting systems is investigated in several materials. \cite{Hebard1990,Yazdani1995,Markovic1998,Bielejec2002,Aubin2006,Giraldo-Gallo2012,Schneider2012} Although magnetic-field-induced SI scaling fails to obtain universal critical resistance,\cite{Yazdani1995} the critical exponents show good agreement with theoretical predictions. Conversely, SI scaling analysis of MoGe\cite{Yazdani1995} thin films shows distinct deviation from standard SI scaling behavior. Magneto-resistance reveals excellent fit to the scaling function for high-temperature isotherms but exhibits remarkable deviation from the scaled curve for low-temperature isotherms. The product of the critical exponents was found to be $\nu \textit{z} \approx$ 1.3 at high temperatures, and this finding closely matches the scaling results for materials such as InO$_{\textit{x}}$ \cite{Hebard1990} that exhibit direct SI transitions. The apparent deviation in the low-temperature isotherms suggests the appearance of the magnetic-field-induced quantum metallic phase. Low-temperature saturated resistance, as well as the deviation from the scaling analysis, supports the existence of the metallic phase. Although scaling analysis is an excellent measure to identify the nature of the transition, no scaling study on the disorder-induced metallic phase has been reported. In this paper, we report scaling analysis for Ta films with a wide range of thicknesses and include the disorder-induced metallic phase in our findings. Using the scaling results for the disorder-induced metallic phase, we investigated the fundamental difference between the magnetic-field-induced and disorder-induced metallic phases.

\begin{figure}[t]
\includegraphics[width=1\columnwidth]{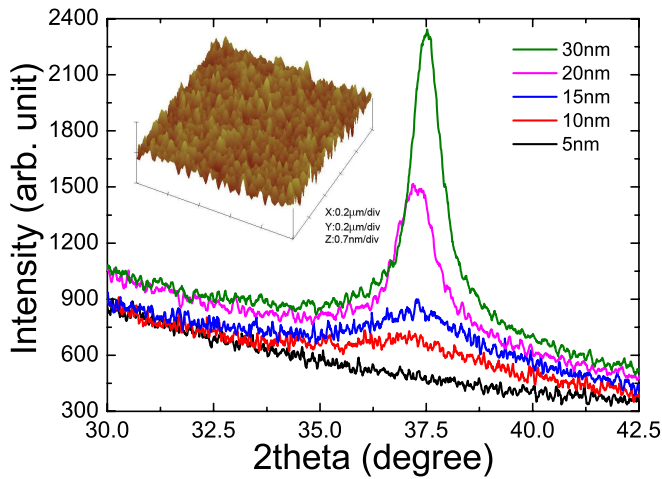}
\caption{\label{fig:xrd} X-ray diffraction patterns of 5-, 10-, 15-, 20-, 30-nm-thick Ta films. Inset: Atomic force microscopy image of the 5-nm Ta thin film. The RMS roughness is approximately 0.1 nm.}
\end{figure}

Ta thin films were fabricated by a DC sputtering technique. The thickness of a sample was determined by a quartz microbalance during sputtering and was confirmed by measurement with an atomic force microscope (AFM) after sputtering. The sample has a Hall-bar shape (1 mm in width and 5 mm in length) for standard four-probe lock-in amplifier measurements. Low-temperature measurements are performed with a home-made cryo-free dilution refrigerator for which the base temperature is 20 mK, and the maximum magnetic field used is 9 T. Ta thin films are then characterized by X-ray diffraction (XRD) and AFM measurements. Figure \ref{fig:xrd} shows the XRD results for Ta thin films with various thicknesses. Ta films with a thickness greater than 5 nm reveal a crystalline peak at $\approx$ 37$^{\circ}$, which is attributed to the local body-centered cubic ($\alpha$-phase) correlation\cite{Read1972}. However, films with a thickness less than 5 nm do not show any peak structures in XRD, which indicates that these thin films are highly amorphous. An additional AFM image of the 5-nm Ta thin film is shown in the inset of Fig. \ref{fig:xrd}. The RMS roughness is about 0.1 nm, which suggests that the film is spatially homogeneous. This roughness value is an order of magnitude smaller than the value for the InO$_{\textit{x}}$ thin films.\cite{Tan2008}

\begin{figure}[bp]
\includegraphics[width=0.8\columnwidth]{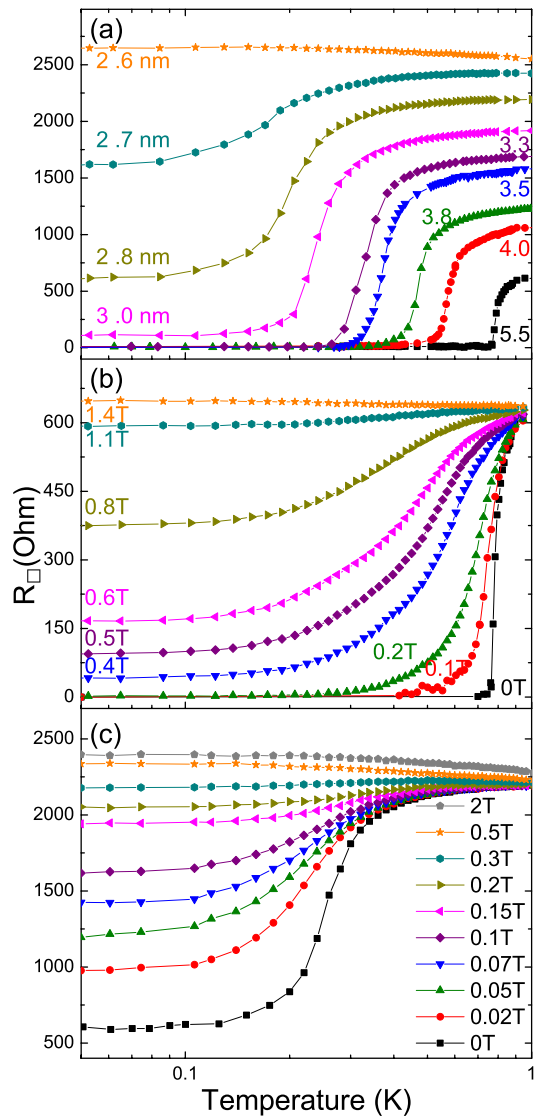}
\caption{\label{fig:RT} (a) Temperature dependence of sheet resistance for Ta thin films of various thicknesses. The film thicknesses (from the bottom upward) are 5.5, 4.0, 3.8, 3.5, 3.3, 3.0, 2.8, 2.7, and 2.6 nm. (b) Temperature dependence of sheet resistance under the indicated magnetic field in the 5.5-nm sample. (c) Temperature dependence of sheet resistance under the indicated magnetic field in the 2.8-nm sample.}
\end{figure}

The sheet resistance, $\textit{R}_{\square}$, for various Ta film thicknesses is plotted as a function of temperature in the absence of a magnetic field, as shown in Fig. \ref{fig:RT}(a). Superconducting thin films with a thickness between 3.5 nm and 5.5 nm show an abrupt resistance drop to zero at critical temperature $\textit{T}_\textit{c}$, and no evidence of re-entrant behavior is found. The critical temperature of superconducting films decreases with decreasing film thickness. This thickness dependence of $\textit{T}_\textit{c}$ is a distinct feature observed in many other amorphous and homogeneous thin films,\cite{Goldman1998} while the $\textit{T}_\textit{c}$ of granular superconducting films is independent of film thickness.\cite{Hsu1993}

We observed the appearance of a disorder-induced phase that cannot be classified as either the superconducting or the insulating phase. For Ta films between 2.7 nm and 3.0 nm, the resistance drop is not as sharp as the decline observed in typical superconducting samples. Compared with superconducting samples, the resistance decreases with decreasing temperature at a somewhat slower rate and is, finally, saturated to the measurable finite value at the low-temperature limit. The broadening of transition is further enhanced, and the saturated final resistances at low temperatures increase with decreasing thickness. This phase can be considered a disorder-induced metallic phase, which is consistent with the previous observation. Finally, further decreasing the thickness to 2.6 nm reveals a negative slope in d$\textit{R}$/d\textit{T}, which indicates that this Ta film is insulating. We find that these characteristic features of the disorder-tuned superconductor-metal-insulator transition are consistent with previous studies.\cite{Li2010} 
Figures \ref{fig:RT}(b) and (c) show the sheet resistance of Ta films with two different thicknesses – 5.5 nm and 2.8 nm, respectively – with various magnetic fields as a function of temperature. At high temperatures where Ta films remained normal state, the magnetic field perpendicular to the samples was set to a target value, and the sheet resistances are measured while the samples cool. We find that the escalation of the magnetic field alters the nature of the transition dramatically. For instance, in the presence of a magnetic field of 0.4 T, the superconducting thin film with a 5.5-nm thickness is transformed to the metallic phase with finite saturated resistance at low temperatures. The field-induced metallic phase was found in a wide range of magnetic fields for which the characteristic features were largely consistent with the previous results reported by Yoon’s group.\cite{Qin2006,Seo2006,Li2010} With an increasing magnetic field, the saturated resistance increases monotonically toward normal-state resistance. The sheet resistance measured at B = 1.4 T exhibits negative d$\textit{R}$/d\textit{T} dependence and thereby demonstrates that the sample becomes an insulator. For the disorder-induced metallic sample, the saturated finite sheet resistance monotonically increases with increasing magnetic field and subsequently enters an insulating phase at a magnetic field higher than 0.5 T, as shown in Fig. 2(c).

\begin{figure}[t]
\includegraphics[width=1\columnwidth]{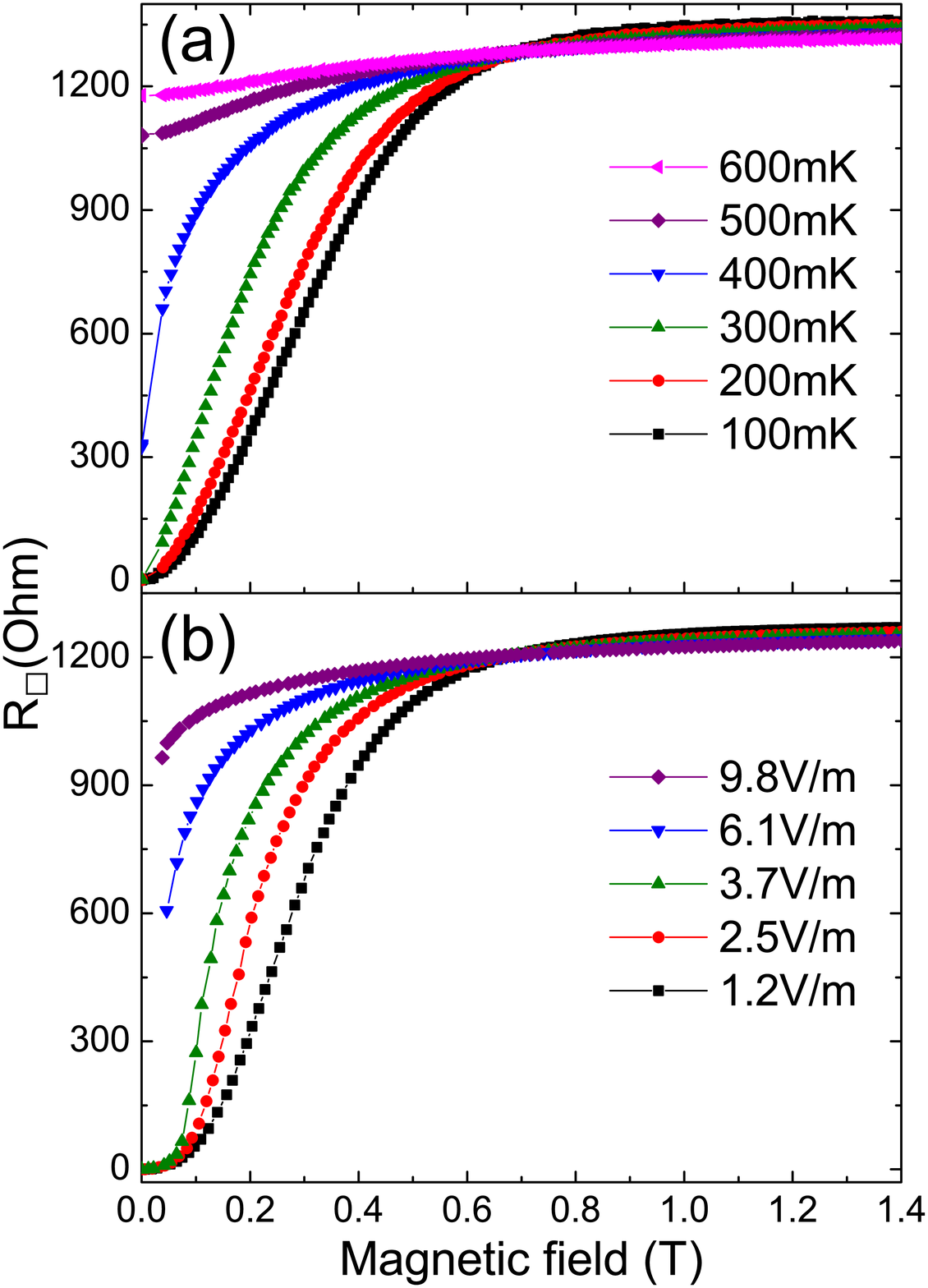}
\caption{\label{fig:BT1} (a) Magnetic field dependence of sheet resistance at indicated temperatures in the superconducting 3.8-nm sample. (b) Magnetic field dependence of sheet resistance with various electric fields at \textit{T} = 50 mK.}
\end{figure}

\begin{figure}[t]
\includegraphics[width=1\columnwidth]{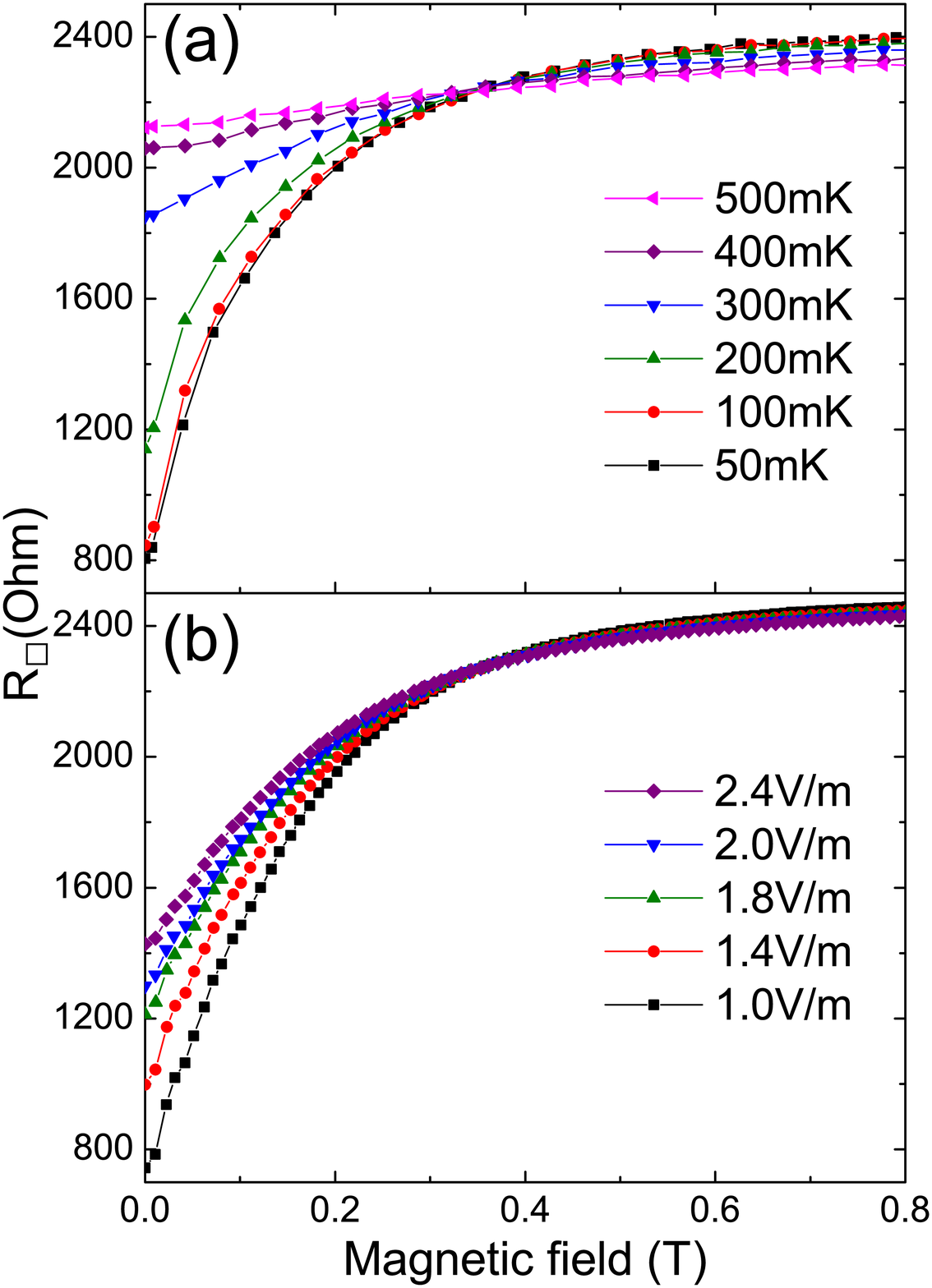}
\caption{\label{fig:BT2} (a) Magnetic field dependence of sheet resistance at indicated temperatures in the metallic 2.8-nm sample. (b) Magnetic field dependence of sheet resistance with various electric fields near the transition at \textit{T} = 50 mK.}
\end{figure}

The magneto-resistance curves for the 3.8-nm-thick superconducting sample measured at various temperatures are shown in Fig. \ref{fig:BT1}(a). The sample was cooled to target temperatures without a magnetic field, and the sheet resistance was measured with an increasing magnetic field. The magneto-resistance isotherms cross at a characteristic value of the magnetic field where the resistance is independent of the temperature. The d$\textit{R}$/d\textit{T} is positive below this critical point and is negative above this point. This characteristic point can be defined as the critical field, $\textit{B}_\textit{c}$, which divides the sample between the insulating phase and the superconducting phase. The disorder-induced metallic samples share essentially identical features with the superconducting samples. The magneto-resistance isotherms for the 2.8-nm-thick metallic sample are plotted in Fig. \ref{fig:BT2}(a). The critical field decreased systematically by by decreasing film thickness, and no negative magneto-resistance region was found in any thickness range, which is a result reported for several thin films.\cite{Steiner2005,Paalanen1992,Sam2004,Baturina2004,Nguyen2009}

Instead of temperature, we measured the magneto-resistance curves for various electric fields at the base temperature. The resistance was measured at a certain target electric field as the magnetic field increased. We observed approximately the same critical field, $\textit{B}_\textit{c}$, in this set of measurements for the fields measured at various temperatures. The magneto-resistance, as shown in Figs. \ref{fig:BT1}(b) and Figs. \ref{fig:BT2}(b), is independent of the electric field at this critical point, $\textit{B}_\textit{c}$. The sheet resistance increases when measured with increasing electric field below $\textit{B}_\textit{c}$ and decreases above this value. 

At the critical point of the magnetic-field-induced SI transition, the correlation length diverges as $\xi \approx \delta^{-\nu}$ , and the characteristic frequency vanishes as $\Omega \approx \xi^{-\textit{z}}$, where $\delta = \vert \textit{B}-\textit{B}_\textit{c} \vert$, $\nu$ is the correlation length exponent, and $\textit{z}$ is the dynamic critical exponent. Since the critical fluctuation is the most important fluctuation near the transition, critical exponents $\nu$ and $\textit{z}$ are independent of materials and microscopic details when the microscopic disorder is sufficiently homogeneous. The scaling behavior of a 2D disordered superconducting system near the critical field can be expressed by \cite{Sondhi1997,Yazdani1995}

\begin{equation}
R_{\square}(B,T,E)=R_cf(\frac{c_a\delta}{T^{1/z\nu}},\frac{c_b\delta}{E^{1/(z+1)\nu}})
\end{equation}

where $\textit{R}_\textit{c}$ = h/4e$^2$ is the universal sheet resistance at the critical field and \textit{f(x)} is the universal scaling function. Exponents $\nu$ and \textit{z} cannot be obtained independently by a single scaling analysis of either electric-field-tuned measurements or temperature-tuned measurements; this allows us to determine only the product value of either $\nu \textit{z}$ or $\nu$(\textit{z}+1), respectively. Apparently, independent determination of critical exponents  $\nu$ and $\textit{z}$ can seemingly be achieved by combining both measurements.\cite{Yazdani1995} 

\begin{figure}[t]
\includegraphics[width=1\columnwidth]{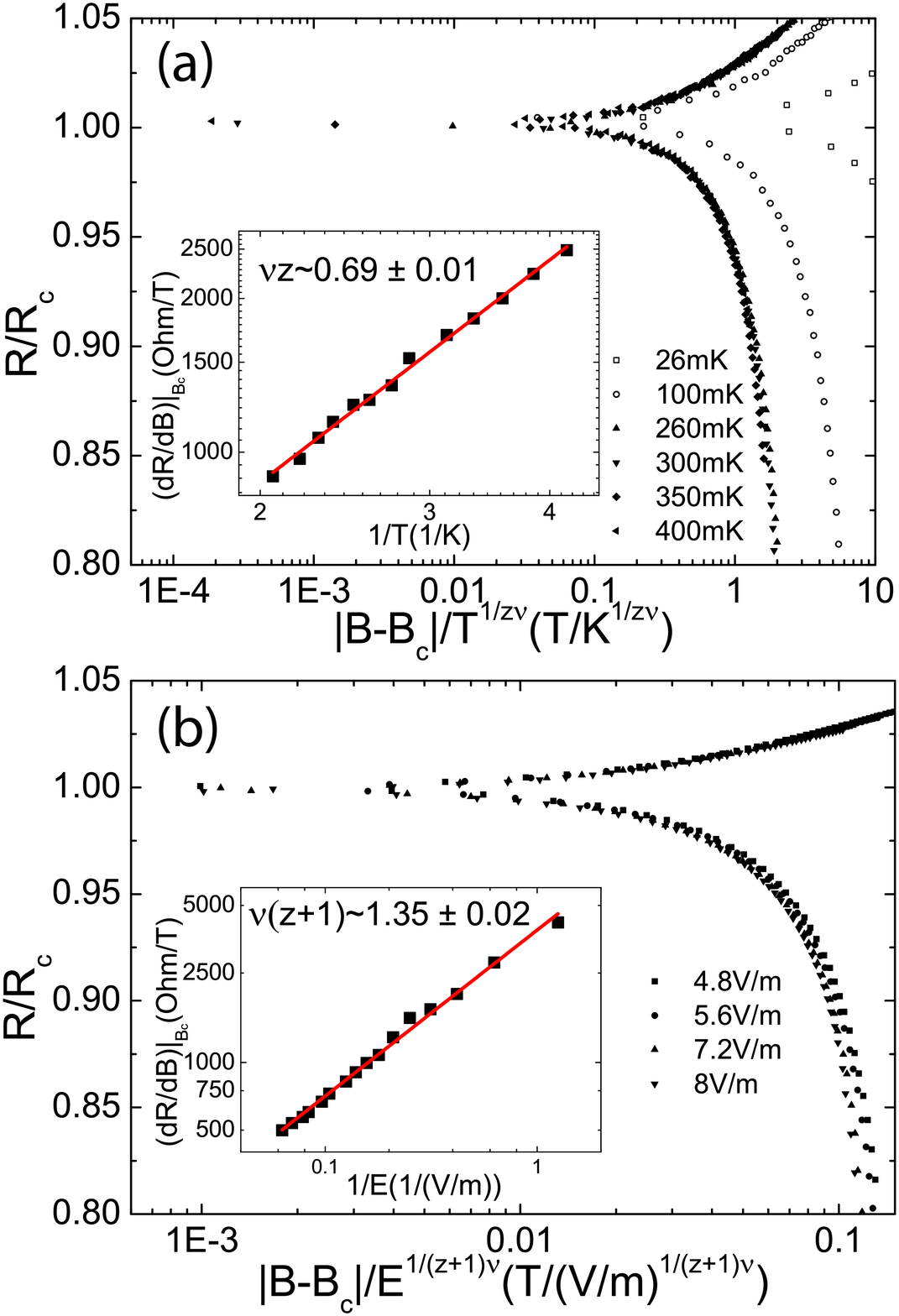}
\caption{\label{fig:ss} (a) Normalized resistance of the superconducting sample as a function of the scaling variable $\vert \textit{B}-\textit{B}_\textit{c} \vert / \textit{T}^{1/\textit{z}\nu}$ at indicated temperatures (for clarity, only six temperatures are shown). Inset: The fitting of a power law to the inverse temperature dependence of d$\textit{R}$/d\textit{B} at $\textit{B}_\textit{c}$. (b) Normalized resistance of the identical superconducting sample as a function of the scaling variable $\vert \textit{B}-\textit{B}_\textit{c} \vert / \textit{E}^{1/(\textit{z}+1)\nu}$ at indicated electric fields (for clarity, only four electric fields are shown). Inset: The fitting of a power law to the inverse electrical field dependence of d$\textit{R}$/d\textit{B} at $\textit{B}_\textit{c}$.}
\end{figure}

We utilized $\textit{B}_\textit{c}$, which is determined from the magneto-resistance isotherms shown in Fig. \ref{fig:BT1}, in the scaling analysis for temperature and the electric field. In Figs. \ref{fig:ss}(a) and (b), the magneto-resistance isotherms of the superconducting sample are plotted as a function of the scaling variable $\frac{\delta}{\textit{T}^{1/\textit{z}\nu}}$ for the temperature scaling analysis and $\frac{\delta}{\textit{E}^{1/(\textit{z}+1)\nu}}$ for the electric field scaling analysis. We determined the product $\nu \textit{z}$ by evaluating the mathematical relation $\frac{d\textit{R}}{d\textit{B}} \vert _{\textit{B}_\textit{c}} \propto \textit{R}_\textit{c} \textit{T }^{-1/\textit{z}\nu} \textit{f }^{\prime}(0)$.\cite{Hebard1990} d$\textit{R}$/d\textit{B} at $\textit{B}_\textit{c}$ is depicted as a function of the inverse temperature in the inset of Fig.\ref{fig:ss}(a), and the product $\nu \textit{z}$ is determined by evaluating the inverse slope of the log-log plot of $\frac{d\textit{R}}{d\textit{B}} \vert _{\textit{B}_\textit{c}}$ vs. 1/\textit{T}. The critical exponent product is found to be 0.69 $\pm$ 0.01, which appears inconsistent with theoretical predictions for which z is expected to be 1 in a bosonic system with coulomb interactions, while $\nu$ is expected to be greater than 1 in a two-dimensional transition tuned by the disorder strength.\cite{Fisher1990,Fisher1990a} Without applying any fitting parameter, the scaling function can be directly tested by substituting $\textit{B}_\textit{c}$ and $\nu \textit{z}$ into the function. The resistance isotherms measured at temperatures higher than 0.2 K collapse onto a single curve, as shown in Fig. \ref{fig:ss}(a), which indicates that the scaling function is well obeyed. However, the resistance isotherms measured at low temperatures deviate from the scaled curve of the high-temperature isotherms, which suggests that the transition at low temperatures does not belong to the same universality class as the SI transition. We notice that a similar low-temperature deviation was found in MoGe superconducting films, and this deviation was ascribed to the coupling of the system to a dissipative bath.\cite{Mason1999}

We investigated the consistency of the temperature scaling analysis by introducing a different method to obtain the critical exponents: electric field scaling analysis. The magneto-resistance was measured in samples with various electric fields at \textit{T} = 50 mK, where no apparent change in resistance appears as a function of temperature. We determined the product $\nu$(\textit{z}+1) from the electric field scaling analysis using essentially the same procedure applied to the preceding temperature scaling analysis. The mathematical relation $\frac{d\textit{R}}{d\textit{B}} \vert _{\textit{B}_\textit{c}} \propto \textit{R}_\textit{c} \textit{E }^{-1/(\textit{z}+1)\nu} \textit{f }^{\prime}(0)$ was used to evaluate the product $\nu$(\textit{z}+1).\cite{Yazdani1995} As shown in the inset of Fig. \ref{fig:ss}(b), the inverse slope of $\frac{d\textit{R}}{d\textit{B}} \vert _{\textit{B}_\textit{c}}$ as a function of the inverse electric field (plotted in logarithmic scale) was used to find $\nu$(\textit{z}+1) = 1.35 $\pm$ 0.02. The scaling behavior is very well obeyed for sheet resistance isotherms plotted using the exponent product $\nu$(\textit{z}+1), as shown in Fig. \ref{fig:ss}(b). The critical exponents of the superconducting sample from both independent procedures are determined to be $\nu \approx 0.66 \pm 0.04$ and $\textit{z} \approx 1.04 \pm 0.08$, which indicate that critical exponent $\nu$ is inconsistent with the theoretically predicted value.\cite{Fisher1990} We notice that this result, however, agrees with the case in which disorder is absent. The universality class of the classical 3D XY model – which is equivalent to two-dimensional systems without disorder – leads to $\nu \approx 0.7$. In addition, numerical simulations for the Boson Hubbard model in a two-dimensional system without disorder also reveal a coherence length exponent of 0.7. It was also proposed that the correlation length can be altered by disorder averaging, which might allow for $\nu$ less than 1 even in a disordered system.\cite{Pazmandi1997} We also find that this result is clearly different from the experimental results of InO$_{\textit{x}}$\cite{Hebard1990} and MoGe,\cite{Yazdani1995} but is consistent with the results for a-Bi.\cite{Markovic1998} We speculate that the discrepancies may be explained by the differences in the nature of the disorder between composite materials such as InO$_{\textit{x}}$ and MoGe, and monatomic materials such as a-Bi and Ta. For thin films comprised of monatomic materials, the disorder is solely controlled by film thickness, while disorder is controlled by the composition ratio and the annealing conditions in addition to thickness for composite materials. The abundant microstructures induced by various atomic compositions and annealing processes can augment the random disorder in thin films.

\begin{figure}[t]
\includegraphics[width=1\columnwidth]{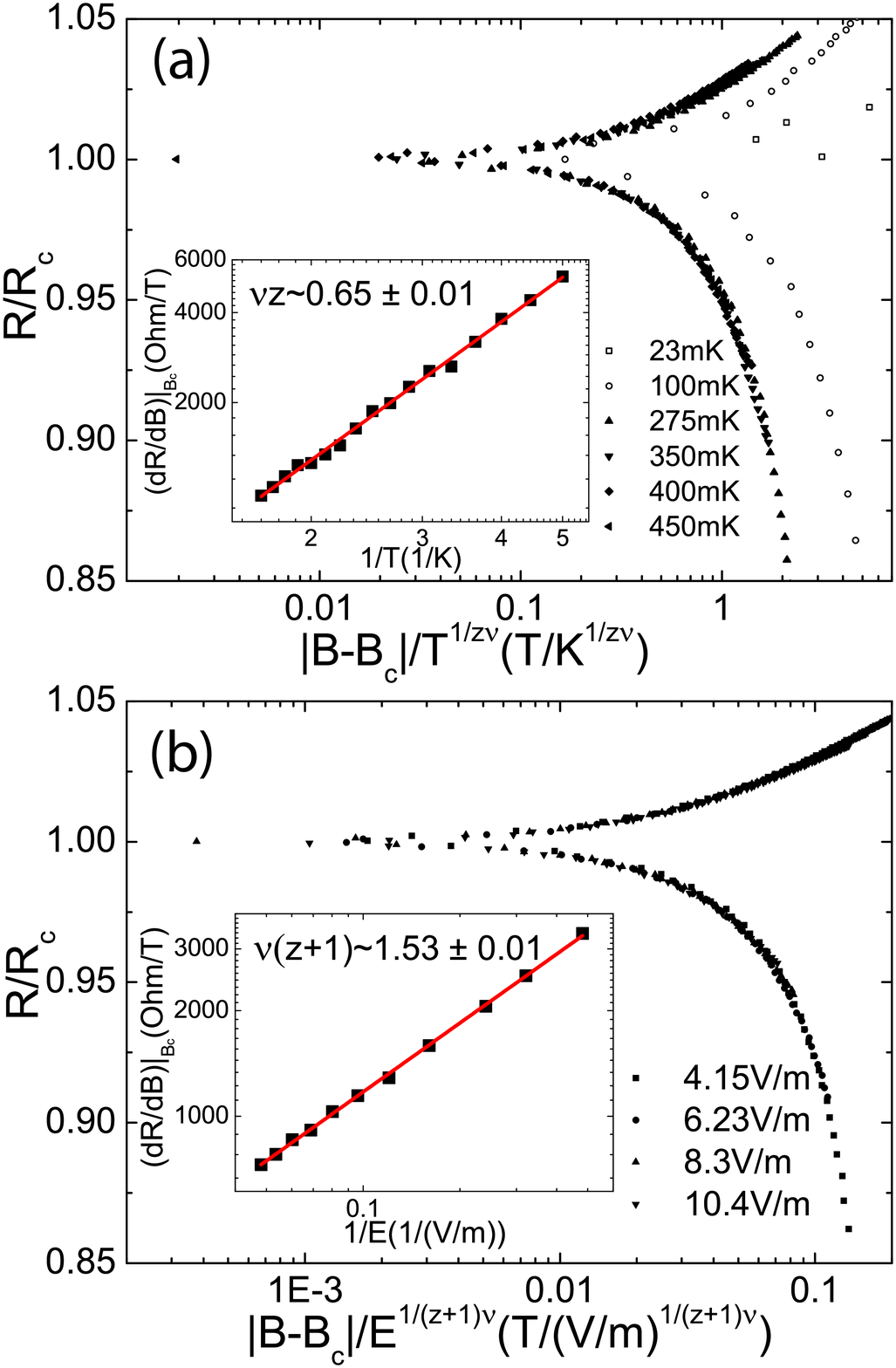}
\caption{\label{fig:ms} (a) Normalized resistance of the metallic sample as a function of the scaling variable $\vert \textit{B}-\textit{B}_\textit{c} \vert / \textit{T}^{1/\textit{z}\nu}$ at indicated temperatures (for clarity, only six temperatures are shown). Inset: The fitting of a power law to the inverse temperature dependence of d$\textit{R}$/d\textit{B} at $\textit{B}_\textit{c}$. (b) Normalized resistance of the identical metallic sample as a function of the scaling variable $\vert \textit{B}-\textit{B}_\textit{c} \vert / \textit{E}^{1/(\textit{z}+1)\nu}$ at indicated electric fields (for clarity, only four electric fields are shown). Inset: The fitting of a power law to the inverse electrical field dependence of d$\textit{R}$/d\textit{B} at $\textit{B}_\textit{c}$.}
\end{figure}

The temperature scaling result for the disorder-induced metallic sample is shown in Fig. \ref{fig:ms}(a), and the electric field scaling is shown in Fig. \ref{fig:ms}(b). We find that the scaling curves are visually well collapsed with the same scaling function used in the superconducting sample. Furthermore, the low-temperature deviation from the main scaling curves in the metallic sample resembles the deviation in the superconducting sample. The critical exponent product $\nu \textit{z}$ 
for the metallic sample is found to be 0.65 $\pm$ 0.01, which is approximately identical to the value obtained in the superconducting sample. However, the product of $\nu$(\textit{z}+1) determined from the electric-field-tuned analysis is found to be about 1.53 $\pm$ 0.01, which represents a clear discrepancy from the superconducting sample. From the combination of the two measurements, we obtained $\nu \approx 0.88 \pm 0.02$ and $\textit{z} \approx 0.74 \pm 0.03$, which differ substantially from theoretical predictions and previous experimental results.

As previously mentioned, in a two-dimensional electron system with long-range Coulomb interaction, the theoretical prediction for critical exponent \textit{z} is unity, which is consistent with experimental results for different transitions, including the metal-insulator transition,\cite{Kravchenko1995,Kravchenko1996} the quantum Hall transition,\cite{Sondhi1997} and the SI transition. The significant discrepancy in the dynamic critical exponent for the disorder-induced metallic sample implies that standard SI scaling analysis cannot be applied to a disorder-induced metallic sample. We believe that the collapse of the scaling curves with certain critical exponents might be coincidental. 

\begin{figure}[t]
\includegraphics[width=1\columnwidth]{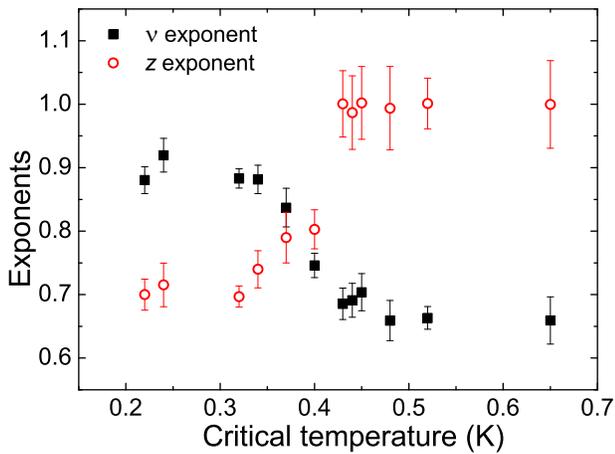}
\caption{\label{fig:pd} temperature dependence of critical exponents ν and z for a wide range of disorders.}
\end{figure}

Figure \ref{fig:pd} depicts critical exponents $\nu$ and $\textit{z}$ as functions of the critical temperature of the SI transition for Ta films of various thicknesses. The Ta samples are superconducting above a certain critical value of the transition temperature and are metallic below that value. Pronounced transition in the critical exponents from metallic to superconducting samples is found as shown in Fig. \ref{fig:pd}. Although the metallic sample may require a new method of analysis to obtain its critical exponents, the clear transition indicates that the metallic sample does not belong to the same universality class as the superconducting samples.

To investigate the scaling analysis for the disorder-induced metallic sample, it is seemingly necessary to find a new scaling analysis, such as metal-insulator (MI) transition scaling.\cite{Das2001} The superconductor-metal transition is associated with the unbinding of quantum dislocation-antidislocation pairs. The film enters a metallic state due to strong gauge-field fluctuations. At the critical point for the MI transition, a MI scaling function is proposed with $ \textit{R}[\frac{\textit{T}^{1/\textit{z}\nu}}{\delta}]^{\nu(\textit{z}+2)} = \textit{f}(\delta/T^{1/\textit{z}\nu})$. With this scaling function, we find that the low-temperature deviation disappears, and scaling curves are well collapsed at all temperature ranges. The scaling function is, however, insensitive to the choice of critical exponents due to the diverging nature at the critical point, which does not allow us to determine these well-defined critical exponents.

A low-temperature deviation from the SI scaling curve is found both in superconducting and metallic samples, although the nature of these two transitions may be fundamentally different. Since the geometry of the film is fixed while the magnetic field increases, the similar deviation suggests that the role of magnetic field is essentially identical in both samples. The increase of magnetic field adds vortices which behave as point disorders and condense at sufficiently high densities, and these vortices thereby drive the system insulating. Because vortices can be added more efficiently at higher temperatures and are strongly frozen at lower temperatures, the microstructure of disorders and phase fluctuations induced by the magnetic field may be temperature dependent. We speculate that the difference in the disorder averaged between high and low temperature may lead to the disagreement in the scaling behavior.

Assuming that the disorder-induced metallic phase arises due to the same mechanism as the magnetic-field-induced metallic phase, one may expect that the SI scaling analysis should be well obeyed only at high temperatures. However, the disorder-induced metallic samples show the evident failure of the SI scaling analysis, even at high temperatures. Accordingly, one could argue that the microscopic mechanism that causes the disorder-induced metallic phase is seemingly different. 

We also hypothesized that the magnetic-field-induced phase transition is also partially dependent on the nature of the disorder. The appearance of negative magneto-resistance was attributed\cite{Ghosal1998,Dubi2006,Dubi2007} to the spatial fluctuations of the superconducting order parameter amplitude, which were more significant when the magnetic field was introduced in highly disordered thin films. We, however, found that Ta thin films exhibited no negative magneto-resistance in any thickness range, which suggests the nature of the disorder in Ta thin films could be different from that of those highly disordered thin films. We speculate that this difference could result from the spatially smooth surface of Ta films, as shown in the AFM images, which is also consistent with results obtained for different morphologies of Bi thin films.\cite{Hollen2011,Hollen2013} Further investigation into the role of different disorders is necessary to understand the interplay between magnetic-field-induced mobile disorders and spatial disorders induced by various microscopic structures in thin films.

In summary, the scaling analysis for Ta thin films with a wide range of disorders is performed. Critical exponents $\nu$ and $\textit{z}$ are obtained independently by performing temperature-tuned and electric-field-tuned scaling analysis. The superconducting samples reveal that the well-obeyed dynamic critical exponent is approximately unity, whereas these samples also show the apparent deviation of critical exponent $\nu$ from the theoretical predictions and from previous experimental results. Additionally, we find that the critical exponents obtained from the disorder-induced metallic samples – which are clearly inconsistent with theoretical predictions – show a marked discrepancy from the exponents obtained for the superconducting samples. The apparent transition between these two types of samples indicates that the disorder-induced metallic film does not belong to the same universality class as two-dimensional Ta superconducting films. Further investigation is necessary to understand the microscopic mechanism underlying this Bose metallic phase intervening superconductor and insulator phase.

\begin{acknowledgments}
The authors acknowledge Jongsoo Yoon for fruitful discussions and his assistance in the early stages of this work. And this work is supported by the National Research Foundation of Korea through the Creative Research Initiatives. 
\end{acknowledgments}

\bibliography{scaling}

%merlin.mbs apsrev4-1.bst 2010-07-25 4.21a (PWD, AO, DPC) hacked
%Control: key (0)
%Control: author (8) initials jnrlst
%Control: editor formatted (1) identically to author
%Control: production of article title (-1) disabled
%Control: page (0) single
%Control: year (1) truncated
%Control: production of eprint (0) enabled
\begin{thebibliography}{40}%
\makeatletter
\providecommand \@ifxundefined [1]{%
 \@ifx{#1\undefined}
}%
\providecommand \@ifnum [1]{%
 \ifnum #1\expandafter \@firstoftwo
 \else \expandafter \@secondoftwo
 \fi
}%
\providecommand \@ifx [1]{%
 \ifx #1\expandafter \@firstoftwo
 \else \expandafter \@secondoftwo
 \fi
}%
\providecommand \natexlab [1]{#1}%
\providecommand \enquote  [1]{``#1''}%
\providecommand \bibnamefont  [1]{#1}%
\providecommand \bibfnamefont [1]{#1}%
\providecommand \citenamefont [1]{#1}%
\providecommand \href@noop [0]{\@secondoftwo}%
\providecommand \href [0]{\begingroup \@sanitize@url \@href}%
\providecommand \@href[1]{\@@startlink{#1}\@@href}%
\providecommand \@@href[1]{\endgroup#1\@@endlink}%
\providecommand \@sanitize@url [0]{\catcode `\\12\catcode `\$12\catcode
  `\&12\catcode `\#12\catcode `\^12\catcode `\_12\catcode `\%12\relax}%
\providecommand \@@startlink[1]{}%
\providecommand \@@endlink[0]{}%
\providecommand \url  [0]{\begingroup\@sanitize@url \@url }%
\providecommand \@url [1]{\endgroup\@href {#1}{\urlprefix }}%
\providecommand \urlprefix  [0]{URL }%
\providecommand \Eprint [0]{\href }%
\providecommand \doibase [0]{http://dx.doi.org/}%
\providecommand \selectlanguage [0]{\@gobble}%
\providecommand \bibinfo  [0]{\@secondoftwo}%
\providecommand \bibfield  [0]{\@secondoftwo}%
\providecommand \translation [1]{[#1]}%
\providecommand \BibitemOpen [0]{}%
\providecommand \bibitemStop [0]{}%
\providecommand \bibitemNoStop [0]{.\EOS\space}%
\providecommand \EOS [0]{\spacefactor3000\relax}%
\providecommand \BibitemShut  [1]{\csname bibitem#1\endcsname}%
\let\auto@bib@innerbib\@empty
%</preamble>
\bibitem [{\citenamefont {Sondhi}\ \emph {et~al.}(1997)\citenamefont {Sondhi},
  \citenamefont {Girvin}, \citenamefont {Carini},\ and\ \citenamefont
  {Shahar}}]{Sondhi1997}%
  \BibitemOpen
  \bibfield  {author} {\bibinfo {author} {\bibfnamefont {S.~L.}\ \bibnamefont
  {Sondhi}}, \bibinfo {author} {\bibfnamefont {S.~M.}\ \bibnamefont {Girvin}},
  \bibinfo {author} {\bibfnamefont {J.~P.}\ \bibnamefont {Carini}}, \ and\
  \bibinfo {author} {\bibfnamefont {D.}~\bibnamefont {Shahar}},\ }\href
  {\doibase 10.1103/RevModPhys.69.315} {\bibfield  {journal} {\bibinfo
  {journal} {Rev. Mod. Phys.}\ }\textbf {\bibinfo {volume} {69}},\ \bibinfo
  {pages} {315} (\bibinfo {year} {1997})}\BibitemShut {NoStop}%
\bibitem [{\citenamefont {Hsu}\ \emph {et~al.}(1995)\citenamefont {Hsu},
  \citenamefont {Chervenak},\ and\ \citenamefont {Valles}}]{Hsu1995}%
  \BibitemOpen
  \bibfield  {author} {\bibinfo {author} {\bibfnamefont {S.-Y.}\ \bibnamefont
  {Hsu}}, \bibinfo {author} {\bibfnamefont {J.~A.}\ \bibnamefont {Chervenak}},
  \ and\ \bibinfo {author} {\bibfnamefont {J.~M.}\ \bibnamefont {Valles},
  \bibfnamefont {Jr.}},\ }\href {\doibase 10.1103/PhysRevLett.75.132}
  {\bibfield  {journal} {\bibinfo  {journal} {Phys. Rev. Lett.}\ }\textbf
  {\bibinfo {volume} {75}},\ \bibinfo {pages} {132} (\bibinfo {year}
  {1995})}\BibitemShut {NoStop}%
\bibitem [{\citenamefont {Hebard}\ and\ \citenamefont
  {Paalanen}(1990)}]{Hebard1990}%
  \BibitemOpen
  \bibfield  {author} {\bibinfo {author} {\bibfnamefont {A.~F.}\ \bibnamefont
  {Hebard}}\ and\ \bibinfo {author} {\bibfnamefont {M.~A.}\ \bibnamefont
  {Paalanen}},\ }\href {\doibase 10.1103/PhysRevLett.65.927} {\bibfield
  {journal} {\bibinfo  {journal} {Phys. Rev. Lett.}\ }\textbf {\bibinfo
  {volume} {65}},\ \bibinfo {pages} {927} (\bibinfo {year} {1990})}\BibitemShut
  {NoStop}%
\bibitem [{\citenamefont {Yazdani}\ and\ \citenamefont
  {Kapitulnik}(1995)}]{Yazdani1995}%
  \BibitemOpen
  \bibfield  {author} {\bibinfo {author} {\bibfnamefont {A.}~\bibnamefont
  {Yazdani}}\ and\ \bibinfo {author} {\bibfnamefont {A.}~\bibnamefont
  {Kapitulnik}},\ }\href {\doibase 10.1103/PhysRevLett.74.3037} {\bibfield
  {journal} {\bibinfo  {journal} {Phys. Rev. Lett.}\ }\textbf {\bibinfo
  {volume} {74}},\ \bibinfo {pages} {3037} (\bibinfo {year}
  {1995})}\BibitemShut {NoStop}%
\bibitem [{\citenamefont {Markovic}\ \emph {et~al.}(1998)\citenamefont
  {Markovic}, \citenamefont {Christiansen},\ and\ \citenamefont
  {Goldman}}]{Markovic1998}%
  \BibitemOpen
  \bibfield  {author} {\bibinfo {author} {\bibfnamefont {N.}~\bibnamefont
  {Markovic}}, \bibinfo {author} {\bibfnamefont {C.}~\bibnamefont
  {Christiansen}}, \ and\ \bibinfo {author} {\bibfnamefont {A.~M.}\
  \bibnamefont {Goldman}},\ }\href {\doibase 10.1103/PhysRevLett.81.5217}
  {\bibfield  {journal} {\bibinfo  {journal} {Phys. Rev. Lett.}\ }\textbf
  {\bibinfo {volume} {81}},\ \bibinfo {pages} {5217} (\bibinfo {year}
  {1998})}\BibitemShut {NoStop}%
\bibitem [{\citenamefont {Fisher}(1990)}]{Fisher1990}%
  \BibitemOpen
  \bibfield  {author} {\bibinfo {author} {\bibfnamefont {M.~P.~A.}\
  \bibnamefont {Fisher}},\ }\href {\doibase 10.1103/PhysRevLett.65.923}
  {\bibfield  {journal} {\bibinfo  {journal} {Phys. Rev. Lett.}\ }\textbf
  {\bibinfo {volume} {65}},\ \bibinfo {pages} {923} (\bibinfo {year}
  {1990})}\BibitemShut {NoStop}%
\bibitem [{\citenamefont {Fisher}\ \emph {et~al.}(1990)\citenamefont {Fisher},
  \citenamefont {Grinstein},\ and\ \citenamefont {Girvin}}]{Fisher1990a}%
  \BibitemOpen
  \bibfield  {author} {\bibinfo {author} {\bibfnamefont {M.~P.~A.}\
  \bibnamefont {Fisher}}, \bibinfo {author} {\bibfnamefont {G.}~\bibnamefont
  {Grinstein}}, \ and\ \bibinfo {author} {\bibfnamefont {S.~M.}\ \bibnamefont
  {Girvin}},\ }\href {\doibase 10.1103/PhysRevLett.64.587} {\bibfield
  {journal} {\bibinfo  {journal} {Phys. Rev. Lett.}\ }\textbf {\bibinfo
  {volume} {64}},\ \bibinfo {pages} {587} (\bibinfo {year} {1990})}\BibitemShut
  {NoStop}%
\bibitem [{\citenamefont {Abrahams}\ \emph {et~al.}(1979)\citenamefont
  {Abrahams}, \citenamefont {Anderson}, \citenamefont {Licciardello},\ and\
  \citenamefont {Ramakrishnan}}]{Abrahams1979}%
  \BibitemOpen
  \bibfield  {author} {\bibinfo {author} {\bibfnamefont {E.}~\bibnamefont
  {Abrahams}}, \bibinfo {author} {\bibfnamefont {P.~W.}\ \bibnamefont
  {Anderson}}, \bibinfo {author} {\bibfnamefont {D.~C.}\ \bibnamefont
  {Licciardello}}, \ and\ \bibinfo {author} {\bibfnamefont {T.~V.}\
  \bibnamefont {Ramakrishnan}},\ }\href {\doibase 10.1103/PhysRevLett.42.673}
  {\bibfield  {journal} {\bibinfo  {journal} {Phys. Rev. Lett.}\ }\textbf
  {\bibinfo {volume} {42}},\ \bibinfo {pages} {673} (\bibinfo {year}
  {1979})}\BibitemShut {NoStop}%
\bibitem [{\citenamefont {Qin}\ \emph {et~al.}(2006)\citenamefont {Qin},
  \citenamefont {Vicente},\ and\ \citenamefont {Yoon}}]{Qin2006}%
  \BibitemOpen
  \bibfield  {author} {\bibinfo {author} {\bibfnamefont {Y.}~\bibnamefont
  {Qin}}, \bibinfo {author} {\bibfnamefont {C.~L.}\ \bibnamefont {Vicente}}, \
  and\ \bibinfo {author} {\bibfnamefont {J.}~\bibnamefont {Yoon}},\ }\href
  {\doibase 10.1103/PhysRevB.73.100505} {\bibfield  {journal} {\bibinfo
  {journal} {Phys. Rev. B}\ }\textbf {\bibinfo {volume} {73}},\ \bibinfo
  {pages} {100505} (\bibinfo {year} {2006})}\BibitemShut {NoStop}%
\bibitem [{\citenamefont {Seo}\ \emph {et~al.}(2006)\citenamefont {Seo},
  \citenamefont {Qin}, \citenamefont {Vicente}, \citenamefont {Choi},\ and\
  \citenamefont {Yoon}}]{Seo2006}%
  \BibitemOpen
  \bibfield  {author} {\bibinfo {author} {\bibfnamefont {Y.}~\bibnamefont
  {Seo}}, \bibinfo {author} {\bibfnamefont {Y.}~\bibnamefont {Qin}}, \bibinfo
  {author} {\bibfnamefont {C.~L.}\ \bibnamefont {Vicente}}, \bibinfo {author}
  {\bibfnamefont {K.~S.}\ \bibnamefont {Choi}}, \ and\ \bibinfo {author}
  {\bibfnamefont {J.}~\bibnamefont {Yoon}},\ }\href {\doibase
  10.1103/PhysRevLett.97.057005} {\bibfield  {journal} {\bibinfo  {journal}
  {Phys. Rev. Lett.}\ }\textbf {\bibinfo {volume} {97}},\ \bibinfo {pages}
  {057005} (\bibinfo {year} {2006})}\BibitemShut {NoStop}%
\bibitem [{\citenamefont {Li}\ \emph {et~al.}(2010)\citenamefont {Li},
  \citenamefont {Vicente},\ and\ \citenamefont {Yoon}}]{Li2010}%
  \BibitemOpen
  \bibfield  {author} {\bibinfo {author} {\bibfnamefont {Y.}~\bibnamefont
  {Li}}, \bibinfo {author} {\bibfnamefont {C.~L.}\ \bibnamefont {Vicente}}, \
  and\ \bibinfo {author} {\bibfnamefont {J.}~\bibnamefont {Yoon}},\ }\href
  {\doibase 10.1103/PhysRevB.81.020505} {\bibfield  {journal} {\bibinfo
  {journal} {Phys. Rev. B}\ }\textbf {\bibinfo {volume} {81}},\ \bibinfo
  {pages} {020505} (\bibinfo {year} {2010})}\BibitemShut {NoStop}%
\bibitem [{\citenamefont {Ephron}\ \emph {et~al.}(1996)\citenamefont {Ephron},
  \citenamefont {Yazdani}, \citenamefont {Kapitulnik},\ and\ \citenamefont
  {Beasley}}]{Ephron1996}%
  \BibitemOpen
  \bibfield  {author} {\bibinfo {author} {\bibfnamefont {D.}~\bibnamefont
  {Ephron}}, \bibinfo {author} {\bibfnamefont {A.}~\bibnamefont {Yazdani}},
  \bibinfo {author} {\bibfnamefont {A.}~\bibnamefont {Kapitulnik}}, \ and\
  \bibinfo {author} {\bibfnamefont {M.~R.}\ \bibnamefont {Beasley}},\ }\href
  {\doibase 10.1103/PhysRevLett.76.1529} {\bibfield  {journal} {\bibinfo
  {journal} {Phys. Rev. Lett.}\ }\textbf {\bibinfo {volume} {76}},\ \bibinfo
  {pages} {1529} (\bibinfo {year} {1996})}\BibitemShut {NoStop}%
\bibitem [{\citenamefont {Mason}\ and\ \citenamefont
  {Kapitulnik}(1999)}]{Mason1999}%
  \BibitemOpen
  \bibfield  {author} {\bibinfo {author} {\bibfnamefont {N.}~\bibnamefont
  {Mason}}\ and\ \bibinfo {author} {\bibfnamefont {A.}~\bibnamefont
  {Kapitulnik}},\ }\href {\doibase 10.1103/PhysRevLett.82.5341} {\bibfield
  {journal} {\bibinfo  {journal} {Phys. Rev. Lett.}\ }\textbf {\bibinfo
  {volume} {82}},\ \bibinfo {pages} {5341} (\bibinfo {year}
  {1999})}\BibitemShut {NoStop}%
\bibitem [{\citenamefont {Das}\ and\ \citenamefont {Doniach}(1999)}]{Das1999}%
  \BibitemOpen
  \bibfield  {author} {\bibinfo {author} {\bibfnamefont {D.}~\bibnamefont
  {Das}}\ and\ \bibinfo {author} {\bibfnamefont {S.}~\bibnamefont {Doniach}},\
  }\href {\doibase 10.1103/PhysRevB.60.1261} {\bibfield  {journal} {\bibinfo
  {journal} {Phys. Rev. B}\ }\textbf {\bibinfo {volume} {60}},\ \bibinfo
  {pages} {1261} (\bibinfo {year} {1999})}\BibitemShut {NoStop}%
\bibitem [{\citenamefont {Das}\ and\ \citenamefont {Doniach}(2001)}]{Das2001}%
  \BibitemOpen
  \bibfield  {author} {\bibinfo {author} {\bibfnamefont {D.}~\bibnamefont
  {Das}}\ and\ \bibinfo {author} {\bibfnamefont {S.}~\bibnamefont {Doniach}},\
  }\href {\doibase 10.1103/PhysRevB.64.134511} {\bibfield  {journal} {\bibinfo
  {journal} {Phys. Rev. B}\ }\textbf {\bibinfo {volume} {64}},\ \bibinfo
  {pages} {134511} (\bibinfo {year} {2001})}\BibitemShut {NoStop}%
\bibitem [{\citenamefont {Spivak}\ \emph {et~al.}(2001)\citenamefont {Spivak},
  \citenamefont {Zyuzin},\ and\ \citenamefont {Hruska}}]{Spivak2001}%
  \BibitemOpen
  \bibfield  {author} {\bibinfo {author} {\bibfnamefont {B.}~\bibnamefont
  {Spivak}}, \bibinfo {author} {\bibfnamefont {A.}~\bibnamefont {Zyuzin}}, \
  and\ \bibinfo {author} {\bibfnamefont {M.}~\bibnamefont {Hruska}},\ }\href
  {\doibase 10.1103/PhysRevB.64.132502} {\bibfield  {journal} {\bibinfo
  {journal} {Phys. Rev. B}\ }\textbf {\bibinfo {volume} {64}},\ \bibinfo
  {pages} {132502} (\bibinfo {year} {2001})}\BibitemShut {NoStop}%
\bibitem [{\citenamefont {Dalidovich}\ and\ \citenamefont
  {Phillips}(2002)}]{Dalidovich2002}%
  \BibitemOpen
  \bibfield  {author} {\bibinfo {author} {\bibfnamefont {D.}~\bibnamefont
  {Dalidovich}}\ and\ \bibinfo {author} {\bibfnamefont {P.}~\bibnamefont
  {Phillips}},\ }\href {\doibase 10.1103/PhysRevLett.89.027001} {\bibfield
  {journal} {\bibinfo  {journal} {Phys. Rev. Lett.}\ }\textbf {\bibinfo
  {volume} {89}},\ \bibinfo {pages} {027001} (\bibinfo {year}
  {2002})}\BibitemShut {NoStop}%
\bibitem [{\citenamefont {Galitski}\ \emph {et~al.}(2005)\citenamefont
  {Galitski}, \citenamefont {Refael}, \citenamefont {Fisher},\ and\
  \citenamefont {Senthil}}]{Galitski2005}%
  \BibitemOpen
  \bibfield  {author} {\bibinfo {author} {\bibfnamefont {V.~M.}\ \bibnamefont
  {Galitski}}, \bibinfo {author} {\bibfnamefont {G.}~\bibnamefont {Refael}},
  \bibinfo {author} {\bibfnamefont {M.~P.~A.}\ \bibnamefont {Fisher}}, \ and\
  \bibinfo {author} {\bibfnamefont {T.}~\bibnamefont {Senthil}},\ }\href
  {\doibase 10.1103/PhysRevLett.95.077002} {\bibfield  {journal} {\bibinfo
  {journal} {Phys. Rev. Lett.}\ }\textbf {\bibinfo {volume} {95}},\ \bibinfo
  {pages} {077002} (\bibinfo {year} {2005})}\BibitemShut {NoStop}%
\bibitem [{\citenamefont {Kapitulnik}\ \emph {et~al.}(2001)\citenamefont
  {Kapitulnik}, \citenamefont {Mason}, \citenamefont {Kivelson},\ and\
  \citenamefont {Chakravarty}}]{Kapitulnik2001}%
  \BibitemOpen
  \bibfield  {author} {\bibinfo {author} {\bibfnamefont {A.}~\bibnamefont
  {Kapitulnik}}, \bibinfo {author} {\bibfnamefont {N.}~\bibnamefont {Mason}},
  \bibinfo {author} {\bibfnamefont {S.~A.}\ \bibnamefont {Kivelson}}, \ and\
  \bibinfo {author} {\bibfnamefont {S.}~\bibnamefont {Chakravarty}},\ }\href
  {\doibase 10.1103/PhysRevB.63.125322} {\bibfield  {journal} {\bibinfo
  {journal} {Phys. Rev. B}\ }\textbf {\bibinfo {volume} {63}},\ \bibinfo
  {pages} {125322} (\bibinfo {year} {2001})}\BibitemShut {NoStop}%
\bibitem [{\citenamefont {Bielejec}\ and\ \citenamefont
  {Wu}(2002)}]{Bielejec2002}%
  \BibitemOpen
  \bibfield  {author} {\bibinfo {author} {\bibfnamefont {E.}~\bibnamefont
  {Bielejec}}\ and\ \bibinfo {author} {\bibfnamefont {W.}~\bibnamefont {Wu}},\
  }\href {\doibase 10.1103/PhysRevLett.88.206802} {\bibfield  {journal}
  {\bibinfo  {journal} {Phys. Rev. Lett.}\ }\textbf {\bibinfo {volume} {88}},\
  \bibinfo {pages} {206802} (\bibinfo {year} {2002})}\BibitemShut {NoStop}%
\bibitem [{\citenamefont {Aubin}\ \emph {et~al.}(2006)\citenamefont {Aubin},
  \citenamefont {Marrache-Kikuchi}, \citenamefont {Pourret}, \citenamefont
  {Behnia}, \citenamefont {Berg\'e}, \citenamefont {Dumoulin},\ and\
  \citenamefont {Lesueur}}]{Aubin2006}%
  \BibitemOpen
  \bibfield  {author} {\bibinfo {author} {\bibfnamefont {H.}~\bibnamefont
  {Aubin}}, \bibinfo {author} {\bibfnamefont {C.~A.}\ \bibnamefont
  {Marrache-Kikuchi}}, \bibinfo {author} {\bibfnamefont {A.}~\bibnamefont
  {Pourret}}, \bibinfo {author} {\bibfnamefont {K.}~\bibnamefont {Behnia}},
  \bibinfo {author} {\bibfnamefont {L.}~\bibnamefont {Berg\'e}}, \bibinfo
  {author} {\bibfnamefont {L.}~\bibnamefont {Dumoulin}}, \ and\ \bibinfo
  {author} {\bibfnamefont {J.}~\bibnamefont {Lesueur}},\ }\href {\doibase
  10.1103/PhysRevB.73.094521} {\bibfield  {journal} {\bibinfo  {journal} {Phys.
  Rev. B}\ }\textbf {\bibinfo {volume} {73}},\ \bibinfo {pages} {094521}
  (\bibinfo {year} {2006})}\BibitemShut {NoStop}%
\bibitem [{\citenamefont {Giraldo-Gallo}\ \emph {et~al.}(2012)\citenamefont
  {Giraldo-Gallo}, \citenamefont {Lee}, \citenamefont {Zhang}, \citenamefont
  {Kramer}, \citenamefont {Beasley}, \citenamefont {Geballe},\ and\
  \citenamefont {Fisher}}]{Giraldo-Gallo2012}%
  \BibitemOpen
  \bibfield  {author} {\bibinfo {author} {\bibfnamefont {P.}~\bibnamefont
  {Giraldo-Gallo}}, \bibinfo {author} {\bibfnamefont {H.}~\bibnamefont {Lee}},
  \bibinfo {author} {\bibfnamefont {Y.}~\bibnamefont {Zhang}}, \bibinfo
  {author} {\bibfnamefont {M.~J.}\ \bibnamefont {Kramer}}, \bibinfo {author}
  {\bibfnamefont {M.~R.}\ \bibnamefont {Beasley}}, \bibinfo {author}
  {\bibfnamefont {T.~H.}\ \bibnamefont {Geballe}}, \ and\ \bibinfo {author}
  {\bibfnamefont {I.~R.}\ \bibnamefont {Fisher}},\ }\href {\doibase
  10.1103/PhysRevB.85.174503} {\bibfield  {journal} {\bibinfo  {journal} {Phys.
  Rev. B}\ }\textbf {\bibinfo {volume} {85}},\ \bibinfo {pages} {174503}
  (\bibinfo {year} {2012})}\BibitemShut {NoStop}%
\bibitem [{\citenamefont {Schneider}\ \emph {et~al.}(2012)\citenamefont
  {Schneider}, \citenamefont {Zaitsev}, \citenamefont {Fuchs},\ and\
  \citenamefont {v.~L\"ohneysen}}]{Schneider2012}%
  \BibitemOpen
  \bibfield  {author} {\bibinfo {author} {\bibfnamefont {R.}~\bibnamefont
  {Schneider}}, \bibinfo {author} {\bibfnamefont {A.~G.}\ \bibnamefont
  {Zaitsev}}, \bibinfo {author} {\bibfnamefont {D.}~\bibnamefont {Fuchs}}, \
  and\ \bibinfo {author} {\bibfnamefont {H.}~\bibnamefont {v.~L\"ohneysen}},\
  }\href {\doibase 10.1103/PhysRevLett.108.257003} {\bibfield  {journal}
  {\bibinfo  {journal} {Phys. Rev. Lett.}\ }\textbf {\bibinfo {volume} {108}},\
  \bibinfo {pages} {257003} (\bibinfo {year} {2012})}\BibitemShut {NoStop}%
\bibitem [{\citenamefont {Read}\ and\ \citenamefont
  {Hensler}(1972)}]{Read1972}%
  \BibitemOpen
  \bibfield  {author} {\bibinfo {author} {\bibfnamefont {M.~H.}\ \bibnamefont
  {Read}}\ and\ \bibinfo {author} {\bibfnamefont {D.}~\bibnamefont {Hensler}},\
  }\href {\doibase http://dx.doi.org/10.1016/0040-6090(72)90277-5} {\bibfield
  {journal} {\bibinfo  {journal} {Thin Solid Films}\ }\textbf {\bibinfo
  {volume} {10}},\ \bibinfo {pages} {123 } (\bibinfo {year}
  {1972})}\BibitemShut {NoStop}%
\bibitem [{\citenamefont {Tan}\ \emph {et~al.}(2008)\citenamefont {Tan},
  \citenamefont {Parendo},\ and\ \citenamefont {Goldman}}]{Tan2008}%
  \BibitemOpen
  \bibfield  {author} {\bibinfo {author} {\bibfnamefont {K.~H. S.~B.}\
  \bibnamefont {Tan}}, \bibinfo {author} {\bibfnamefont {K.~A.}\ \bibnamefont
  {Parendo}}, \ and\ \bibinfo {author} {\bibfnamefont {A.~M.}\ \bibnamefont
  {Goldman}},\ }\href {\doibase 10.1103/PhysRevB.78.014506} {\bibfield
  {journal} {\bibinfo  {journal} {Phys. Rev. B}\ }\textbf {\bibinfo {volume}
  {78}},\ \bibinfo {pages} {014506} (\bibinfo {year} {2008})}\BibitemShut
  {NoStop}%
\bibitem [{\citenamefont {Goldman}\ and\ \citenamefont
  {Markovic}(1998)}]{Goldman1998}%
  \BibitemOpen
  \bibfield  {author} {\bibinfo {author} {\bibfnamefont {A.~M.}\ \bibnamefont
  {Goldman}}\ and\ \bibinfo {author} {\bibfnamefont {N.}~\bibnamefont
  {Markovic}},\ }\href@noop {} {\bibfield  {journal} {\bibinfo  {journal}
  {Physics Today}\ }\textbf {\bibinfo {volume} {51}},\ \bibinfo {pages} {39}
  (\bibinfo {year} {1998})}\BibitemShut {NoStop}%
\bibitem [{\citenamefont {Hsu}\ and\ \citenamefont {Valles}(1993)}]{Hsu1993}%
  \BibitemOpen
  \bibfield  {author} {\bibinfo {author} {\bibfnamefont {S.-Y.}\ \bibnamefont
  {Hsu}}\ and\ \bibinfo {author} {\bibfnamefont {J.~M.}\ \bibnamefont
  {Valles}},\ }\href {\doibase 10.1103/PhysRevB.48.4164} {\bibfield  {journal}
  {\bibinfo  {journal} {Phys. Rev. B}\ }\textbf {\bibinfo {volume} {48}},\
  \bibinfo {pages} {4164} (\bibinfo {year} {1993})}\BibitemShut {NoStop}%
\bibitem [{\citenamefont {Steiner}\ and\ \citenamefont
  {Kapitulnik}(2005)}]{Steiner2005}%
  \BibitemOpen
  \bibfield  {author} {\bibinfo {author} {\bibfnamefont {M.}~\bibnamefont
  {Steiner}}\ and\ \bibinfo {author} {\bibfnamefont {A.}~\bibnamefont
  {Kapitulnik}},\ }\href {\doibase
  http://dx.doi.org/10.1016/j.physc.2005.02.014} {\bibfield  {journal}
  {\bibinfo  {journal} {Physica C: Superconductivity}\ }\textbf {\bibinfo
  {volume} {422}},\ \bibinfo {pages} {16 } (\bibinfo {year}
  {2005})}\BibitemShut {NoStop}%
\bibitem [{\citenamefont {Paalanen}\ \emph {et~al.}(1992)\citenamefont
  {Paalanen}, \citenamefont {Hebard},\ and\ \citenamefont
  {Ruel}}]{Paalanen1992}%
  \BibitemOpen
  \bibfield  {author} {\bibinfo {author} {\bibfnamefont {M.~A.}\ \bibnamefont
  {Paalanen}}, \bibinfo {author} {\bibfnamefont {A.~F.}\ \bibnamefont
  {Hebard}}, \ and\ \bibinfo {author} {\bibfnamefont {R.~R.}\ \bibnamefont
  {Ruel}},\ }\href {\doibase 10.1103/PhysRevLett.69.1604} {\bibfield  {journal}
  {\bibinfo  {journal} {Phys. Rev. Lett.}\ }\textbf {\bibinfo {volume} {69}},\
  \bibinfo {pages} {1604} (\bibinfo {year} {1992})}\BibitemShut {NoStop}%
\bibitem [{\citenamefont {Sambandamurthy}\ \emph {et~al.}(2004)\citenamefont
  {Sambandamurthy}, \citenamefont {Engel}, \citenamefont {Johansson},\ and\
  \citenamefont {Shahar}}]{Sam2004}%
  \BibitemOpen
  \bibfield  {author} {\bibinfo {author} {\bibfnamefont {G.}~\bibnamefont
  {Sambandamurthy}}, \bibinfo {author} {\bibfnamefont {L.~W.}\ \bibnamefont
  {Engel}}, \bibinfo {author} {\bibfnamefont {A.}~\bibnamefont {Johansson}}, \
  and\ \bibinfo {author} {\bibfnamefont {D.}~\bibnamefont {Shahar}},\ }\href
  {\doibase 10.1103/PhysRevLett.92.107005} {\bibfield  {journal} {\bibinfo
  {journal} {Phys. Rev. Lett.}\ }\textbf {\bibinfo {volume} {92}},\ \bibinfo
  {pages} {107005} (\bibinfo {year} {2004})}\BibitemShut {NoStop}%
\bibitem [{\citenamefont {Baturina}\ \emph {et~al.}(2004)\citenamefont
  {Baturina}, \citenamefont {Islamov}, \citenamefont {Bentner}, \citenamefont
  {Strunk}, \citenamefont {Baklanov},\ and\ \citenamefont
  {Satta}}]{Baturina2004}%
  \BibitemOpen
  \bibfield  {author} {\bibinfo {author} {\bibfnamefont {T.}~\bibnamefont
  {Baturina}}, \bibinfo {author} {\bibfnamefont {D.}~\bibnamefont {Islamov}},
  \bibinfo {author} {\bibfnamefont {J.}~\bibnamefont {Bentner}}, \bibinfo
  {author} {\bibfnamefont {C.}~\bibnamefont {Strunk}}, \bibinfo {author}
  {\bibfnamefont {M.}~\bibnamefont {Baklanov}}, \ and\ \bibinfo {author}
  {\bibfnamefont {A.}~\bibnamefont {Satta}},\ }\href {\doibase
  10.1134/1.1765178} {\bibfield  {journal} {\bibinfo  {journal} {Journal of
  Experimental and Theoretical Physics Letters}\ }\textbf {\bibinfo {volume}
  {79}},\ \bibinfo {pages} {337} (\bibinfo {year} {2004})}\BibitemShut
  {NoStop}%
\bibitem [{\citenamefont {Nguyen}\ \emph {et~al.}(2009)\citenamefont {Nguyen},
  \citenamefont {Hollen}, \citenamefont {Stewart}, \citenamefont {Shainline},
  \citenamefont {Yin}, \citenamefont {Xu},\ and\ \citenamefont
  {Valles}}]{Nguyen2009}%
  \BibitemOpen
  \bibfield  {author} {\bibinfo {author} {\bibfnamefont {H.~Q.}\ \bibnamefont
  {Nguyen}}, \bibinfo {author} {\bibfnamefont {S.~M.}\ \bibnamefont {Hollen}},
  \bibinfo {author} {\bibfnamefont {M.~D.}\ \bibnamefont {Stewart}}, \bibinfo
  {author} {\bibfnamefont {J.}~\bibnamefont {Shainline}}, \bibinfo {author}
  {\bibfnamefont {A.}~\bibnamefont {Yin}}, \bibinfo {author} {\bibfnamefont
  {J.~M.}\ \bibnamefont {Xu}}, \ and\ \bibinfo {author} {\bibfnamefont {J.~M.}\
  \bibnamefont {Valles}},\ }\href {\doibase 10.1103/PhysRevLett.103.157001}
  {\bibfield  {journal} {\bibinfo  {journal} {Phys. Rev. Lett.}\ }\textbf
  {\bibinfo {volume} {103}},\ \bibinfo {pages} {157001} (\bibinfo {year}
  {2009})}\BibitemShut {NoStop}%
\bibitem [{\citenamefont {Pazmandi}\ \emph {et~al.}(1997)\citenamefont
  {Pazmandi}, \citenamefont {Scalettar},\ and\ \citenamefont
  {Zimanyi}}]{Pazmandi1997}%
  \BibitemOpen
  \bibfield  {author} {\bibinfo {author} {\bibfnamefont {F.}~\bibnamefont
  {Pazmandi}}, \bibinfo {author} {\bibfnamefont {R.~T.}\ \bibnamefont
  {Scalettar}}, \ and\ \bibinfo {author} {\bibfnamefont {G.~T.}\ \bibnamefont
  {Zimanyi}},\ }\href {\doibase 10.1103/PhysRevLett.79.5130} {\bibfield
  {journal} {\bibinfo  {journal} {Phys. Rev. Lett.}\ }\textbf {\bibinfo
  {volume} {79}},\ \bibinfo {pages} {5130} (\bibinfo {year}
  {1997})}\BibitemShut {NoStop}%
\bibitem [{\citenamefont {Kravchenko}\ \emph {et~al.}(1995)\citenamefont
  {Kravchenko}, \citenamefont {Mason}, \citenamefont {Bowker}, \citenamefont
  {Furneaux}, \citenamefont {Pudalov},\ and\ \citenamefont
  {D'Iorio}}]{Kravchenko1995}%
  \BibitemOpen
  \bibfield  {author} {\bibinfo {author} {\bibfnamefont {S.~V.}\ \bibnamefont
  {Kravchenko}}, \bibinfo {author} {\bibfnamefont {W.~E.}\ \bibnamefont
  {Mason}}, \bibinfo {author} {\bibfnamefont {G.~E.}\ \bibnamefont {Bowker}},
  \bibinfo {author} {\bibfnamefont {J.~E.}\ \bibnamefont {Furneaux}}, \bibinfo
  {author} {\bibfnamefont {V.~M.}\ \bibnamefont {Pudalov}}, \ and\ \bibinfo
  {author} {\bibfnamefont {M.}~\bibnamefont {D'Iorio}},\ }\href {\doibase
  10.1103/PhysRevB.51.7038} {\bibfield  {journal} {\bibinfo  {journal} {Phys.
  Rev. B}\ }\textbf {\bibinfo {volume} {51}},\ \bibinfo {pages} {7038}
  (\bibinfo {year} {1995})}\BibitemShut {NoStop}%
\bibitem [{\citenamefont {Kravchenko}\ \emph {et~al.}(1996)\citenamefont
  {Kravchenko}, \citenamefont {Simonian}, \citenamefont {Sarachik},
  \citenamefont {Mason},\ and\ \citenamefont {Furneaux}}]{Kravchenko1996}%
  \BibitemOpen
  \bibfield  {author} {\bibinfo {author} {\bibfnamefont {S.~V.}\ \bibnamefont
  {Kravchenko}}, \bibinfo {author} {\bibfnamefont {D.}~\bibnamefont
  {Simonian}}, \bibinfo {author} {\bibfnamefont {M.~P.}\ \bibnamefont
  {Sarachik}}, \bibinfo {author} {\bibfnamefont {W.}~\bibnamefont {Mason}}, \
  and\ \bibinfo {author} {\bibfnamefont {J.~E.}\ \bibnamefont {Furneaux}},\
  }\href {\doibase 10.1103/PhysRevLett.77.4938} {\bibfield  {journal} {\bibinfo
   {journal} {Phys. Rev. Lett.}\ }\textbf {\bibinfo {volume} {77}},\ \bibinfo
  {pages} {4938} (\bibinfo {year} {1996})}\BibitemShut {NoStop}%
\bibitem [{\citenamefont {Ghosal}\ \emph {et~al.}(1998)\citenamefont {Ghosal},
  \citenamefont {Randeria},\ and\ \citenamefont {Trivedi}}]{Ghosal1998}%
  \BibitemOpen
  \bibfield  {author} {\bibinfo {author} {\bibfnamefont {A.}~\bibnamefont
  {Ghosal}}, \bibinfo {author} {\bibfnamefont {M.}~\bibnamefont {Randeria}}, \
  and\ \bibinfo {author} {\bibfnamefont {N.}~\bibnamefont {Trivedi}},\ }\href
  {\doibase 10.1103/PhysRevLett.81.3940} {\bibfield  {journal} {\bibinfo
  {journal} {Phys. Rev. Lett.}\ }\textbf {\bibinfo {volume} {81}},\ \bibinfo
  {pages} {3940} (\bibinfo {year} {1998})}\BibitemShut {NoStop}%
\bibitem [{\citenamefont {Dubi}\ \emph {et~al.}(2006)\citenamefont {Dubi},
  \citenamefont {Meir},\ and\ \citenamefont {Avishai}}]{Dubi2006}%
  \BibitemOpen
  \bibfield  {author} {\bibinfo {author} {\bibfnamefont {Y.}~\bibnamefont
  {Dubi}}, \bibinfo {author} {\bibfnamefont {Y.}~\bibnamefont {Meir}}, \ and\
  \bibinfo {author} {\bibfnamefont {Y.}~\bibnamefont {Avishai}},\ }\href
  {\doibase 10.1103/PhysRevB.73.054509} {\bibfield  {journal} {\bibinfo
  {journal} {Phys. Rev. B}\ }\textbf {\bibinfo {volume} {73}},\ \bibinfo
  {pages} {054509} (\bibinfo {year} {2006})}\BibitemShut {NoStop}%
\bibitem [{\citenamefont {Dubi}\ \emph {et~al.}(2007)\citenamefont {Dubi},
  \citenamefont {Meir},\ and\ \citenamefont {Avishai}}]{Dubi2007}%
  \BibitemOpen
  \bibfield  {author} {\bibinfo {author} {\bibfnamefont {Y.}~\bibnamefont
  {Dubi}}, \bibinfo {author} {\bibfnamefont {Y.}~\bibnamefont {Meir}}, \ and\
  \bibinfo {author} {\bibfnamefont {Y.}~\bibnamefont {Avishai}},\ }\href@noop
  {} {\bibfield  {journal} {\bibinfo  {journal} {Nature}\ }\textbf {\bibinfo
  {volume} {449}},\ \bibinfo {pages} {876} (\bibinfo {year}
  {2007})}\BibitemShut {NoStop}%
\bibitem [{\citenamefont {Hollen}\ \emph {et~al.}(2011)\citenamefont {Hollen},
  \citenamefont {Nguyen}, \citenamefont {Rudisaile}, \citenamefont {Stewart},
  \citenamefont {Shainline}, \citenamefont {Xu},\ and\ \citenamefont
  {Valles}}]{Hollen2011}%
  \BibitemOpen
  \bibfield  {author} {\bibinfo {author} {\bibfnamefont {S.~M.}\ \bibnamefont
  {Hollen}}, \bibinfo {author} {\bibfnamefont {H.~Q.}\ \bibnamefont {Nguyen}},
  \bibinfo {author} {\bibfnamefont {E.}~\bibnamefont {Rudisaile}}, \bibinfo
  {author} {\bibfnamefont {M.~D.}\ \bibnamefont {Stewart}}, \bibinfo {author}
  {\bibfnamefont {J.}~\bibnamefont {Shainline}}, \bibinfo {author}
  {\bibfnamefont {J.~M.}\ \bibnamefont {Xu}}, \ and\ \bibinfo {author}
  {\bibfnamefont {J.~M.}\ \bibnamefont {Valles}},\ }\href {\doibase
  10.1103/PhysRevB.84.064528} {\bibfield  {journal} {\bibinfo  {journal} {Phys.
  Rev. B}\ }\textbf {\bibinfo {volume} {84}},\ \bibinfo {pages} {064528}
  (\bibinfo {year} {2011})}\BibitemShut {NoStop}%
\bibitem [{\citenamefont {Hollen}\ \emph {et~al.}(2013)\citenamefont {Hollen},
  \citenamefont {Fernandes}, \citenamefont {Xu},\ and\ \citenamefont
  {Valles}}]{Hollen2013}%
  \BibitemOpen
  \bibfield  {author} {\bibinfo {author} {\bibfnamefont {S.~M.}\ \bibnamefont
  {Hollen}}, \bibinfo {author} {\bibfnamefont {G.~E.}\ \bibnamefont
  {Fernandes}}, \bibinfo {author} {\bibfnamefont {J.~M.}\ \bibnamefont {Xu}}, \
  and\ \bibinfo {author} {\bibfnamefont {J.~M.}\ \bibnamefont {Valles}},\
  }\href {\doibase 10.1103/PhysRevB.87.054512} {\bibfield  {journal} {\bibinfo
  {journal} {Phys. Rev. B}\ }\textbf {\bibinfo {volume} {87}},\ \bibinfo
  {pages} {054512} (\bibinfo {year} {2013})}\BibitemShut {NoStop}%
\end{thebibliography}%

\end{document}